\def\rv{{\bf r}}                                   
\def\av{{\bf a}}                                   
\def\vv{{\bf v}}
\def\psum{\sum_{i=0}^{N-1}}
\def\sumv{\psum V_i}
\def\dt{\tau}                                       
\def\veff{{\tilde V}}
\def\sumek{(M/2 \dt) \psum{( \rv_{i+1} - \rv_i )}^2} 
\def\SI{S(\Gamma)}                                     
\def\SIL{S_1(\Gamma)}                                  
\def\SIQ{S_2(\Gamma)}                                  

\magnification=1200
\parskip 2pt
\def\ref#1{[#1]}
\def\msi{\footnote\dag{\rm Partly supported by the U.S. 
  Army Research Office through the Mathematical Sciences Institute of 
  Cornell University}}
\def\add{\footnote\ddag{\rm Mathematical Sciences Institute,
  409 College Av., Ithaca, NY 14850 } }
\medskip
\centerline{\bf Multigrid Solution of a Path Integral Formulation for the
Hydrogen Atom\msi}
\medskip
\centerline{Dov Bai\add}
\medskip
\beginsection 1. Introduction\par

     Most methods for Path Integral Monte Carlo (PIMC) simulations
are inefficient for paths with large number of points. This is due 
mostly to the critical slow down (CSD) for lower dimensions and to
the volume factor \ref{5} for many dimensions. Furthermore, current
applications of PIMC methods for solutions of Schr\"odinger equations for
atomic systems is restricted because of the singularity of the potential
functions. These systems are usually solved by Green's function Monte
Carlo (GFMC) algorithms \ref{2-3}.

    In this work {\it (1)} A path integral formulation appropriate for Coulomb
potentials is presented is Sec. 2. This formulation is based on a 
quadratic polynomial approximation of the classical motion. The resulting
action integral can be expressed as a standard linear expression of 
the action integral with an effective potential. This effective potential
depends on $\tau^2$, where $\tau$ is the mesh size (time step). {\it (2)} A
multigrid algorithm using a ``unigrid'' approach is used for solving
the hydrogen atom (Sec. 3). This algorithm uses a linear interpolation of 
changes to coordinates \ref{5}, rather then constant interpolation
\ref{1}. The use of linear interpolation eliminates the CSD with
simple $V(1,1)$ multigrid cycles. The integrated decorrelation time
$\tau_{int}$ is below 10 for all measured observables. In Sec.4 the
numerical results with the multigrid algorithm are presented. For 
comparison, the same problem is solved with a staging algorithm
\ref{6-11}.


\def\rv{{\bf r}}                                   
\def\av{{\bf a}}                                   
\def\vv{{\bf v}}
\def\psum{\sum_{i=0}^{N-1}}
\def\sumv{\psum V_i}
\def\dt{\tau}                                       
\def\veff{{\tilde V}}
\def\sumek{(M/2 \dt) \psum{( \rv_{i+1} - \rv_i )}^2} 
\def\SI{S(\Gamma)}                                     
\def\SIL{S_1(\Gamma)}                                  
\def\SIQ{S_2(\Gamma)}                                  

\beginsection 2. Path integral formulation for the hydrogen atom \par

In this section $\rv_i, i = 0,\ldots,N-1$ denotes the 3D coordinate 
vector $(x_i, y_i, z_i)$ of the electron, $\tau$ is the meshsize (timestep) of 
the path and $N$ is the number of path points. Since only the ground 
state is considered the path is closed and $\rv_N$ = $\rv_0$. $V_i$ denotes
the potential function at $\rv_i$, and $M$ is the mass of the electron.

The standard 
first order approximation of the action integral for a 
path ${\Gamma}$ is
$$
	\SIL = \dt \sumv + \sumek .           \eqno{(2.1)}
$$
The relative probability for the path is,
$$
	 P(\Gamma) = e^{-\SIL}.
$$
However, this approximation fails for the Coulomb potential
$$
	V(\rv)=-1/r.
$$
The reason is that when the Metropolis \ref{4} algorithm is used to
generate new paths, the path points tend to concentrate near $r=0$, resulting 
in meaningless measurements of observables. This happens because near 
$r=0$, the transition probability from large $r$ is very
close to $1$. Similarly, since the transition probability from $r=0$ to 
large $r$ is very small. Hence a path point can not make the transition 
to larger $r$, once trapped near $r=0$.

In terms of the classical path, the standard approximation (2.1) assumes 
that during the time $\dt$, the particle a) Move along the straight 
line connecting $\rv_i$ and $\rv_{i+1}$, and b) The time that the particle
spends along any portion of that line is proportional to its length.
It is mostly the second assumption that breaks down near the singularity,
since classically if $r_i$ is near 0, the particle spends most of its time 
away from it. 

To make a better approximation of the classical motion, assumption a)
above is preserved, but instead of b) the motion is assumed to be in 
a constant force field. For a motion from $\rv_i$ to $\rv_{i+1}$ the
classical acceleration $\av_i$, is given by
$$
   \av_i = -\hat {\bf u} 
      ( V(\rv_{i+1}) - V(\rv_i) ) / M \vert \rv_{i+1} - \rv_i \vert, \eqno{(2.2)}
$$
where $\hat {\bf u}$ is the unit vector in the direction from $\rv_i$ to
$\rv_{i+1}$. The position as a function of time is,
$$
	\rv(t) = \rv_i +  (\rv_{i+1}-\rv_i)((t-t_i)/\tau) +
	\av_i(t-t_i) ( t-t_i-\tau).    
$$
This classical motion conserves energy. Therefore the total energy $E_i$ 
for the motion $\rv_i \rightarrow \rv_{i+1}$ can be taken as the energy 
at time $t_i$,
$$
	E_i = M{\vv_i}^2/2 - 1 / \rv_i
$$
where $\vv_i$ is 
$$
        \vv_i = ( \rv_{i+1} - \rv_i ) / \tau - \av_i \tau. \eqno{(2.3)}	
$$
Thus, the action integral for a path $\Gamma$ is
$$
	\SIQ = \tau \psum E_i	          		\eqno{(2.4)}	
$$
\beginsection 2.1 Simplified path integral formulation \par

From (2.3), the expression for $\vv_i^2$ contains three terms: ({\it i}) 
${(\rv_{i+1}-\rv_i)}^2/{\dt^2}$. This is the same expression for the
squared velocity that appears in the standard approximation (2.1).
({\it ii}) $\dt^2{\av}^2$. From (2.2), in the limit $\dt\rightarrow 0$,
this term for the Coulomb potential is $O(\dt^2 / r^4 )$. Similarly, in
the same limit the term ({\it iii}) $-2 \av_i (\rv_{i+1} - \rv_i )$ is
$O(1/r^2)$. 

Near $r=0$, the term ({\it ii}) is dominant. For large $r$, the standard
term ({\it i}) is dominant. It is therefore reasonable to approximate
the kinetic energy by preserving only terms ({\it i}) and ({\it ii}),
and omitting ({\it iii}) altogether. The action integral is then,
$$
	\SI = \sumek - \dt \psum 1/r_i + \dt (A/M) \psum \dt^2 / r^4, \eqno{(2.5)}
$$
where $A$ is a positive constant. The last two terms can be 
regarded as a sum over an effective potential
$$
	\veff_i (\rv,\dt) = -1/r_i + A \dt^2 / M r_i^4.  \eqno{(2.6)}
$$

Note that the last term in (2.6) acts as a repulsive 
potential, and thus prevents the path points from falling into $r=0$. For a
fixed $u=A\dt^2/M{r(u)}^4$, $r(u)$ decreases as $\sqrt\dt$. 
With $\veff$ the action integral is 
$$ 
	\SI = \sumek + \dt \psum \veff_i.                \eqno{(2.7)}
$$
This considerably simplifies the Monte Carlo sampling of paths, as
its form is the same as the standard approximation (2.1). 
\nobreak

\def\dt{\tau}                                       
\def\ref#1{[#1]}
\def\gij#1#2{\Gamma_{#1,#2}}
\def\Sij#1#2{S_{#1,#2}}
\def\sumij#1#2{\sum_{i=#1}^{#2}}
\def\ekij#1#2{(M/2 \dt) \sumij{#1}{#2}{( \rv_{i+1} - \rv_i )}^2} 
\def\veff{{\tilde V}}
\def\rv{{\bf r}}                                   
\def\abs#1{{\vert #1 \vert}}
\beginsection 3. Multigrid algorithm \par

To eliminate the correlation between paths, a multigrid algorithm is
used for generating paths. The algorithm in this study implements a
``unigrid'' approach, because of its simplicity. In this approach there 
is only a single grid and coarse level operations are emulated on this grid. 
Such an algorithm in not very effective in reducing the {\it volume factor} 
(see. \ref{5}) for a large number of dimensions. However, it is effective in
eliminating the CSD (see Sec. 4), which is the most important 
factor in efficiency degradation for lower number of dimensions such as the
hydrogen atom. For higher dimensions, a fully multigridded algorithm 
should be used.

The finest level $l=0$ contains the whole set of points 
$i=0, \ldots, N-1$ of the path, corresponding to times $t_i = i\dt$. $N$ is 
assumed to be an integral power of 2. A coarser level $l=1,\ldots,l_c$ 
contains the set of $n_l = N / 2^l$ points $i_l= n2^l$, 
$n = 0, 1, \ldots, n_l$, corresponding to the times $t_{i_l} = t_{i_l} \dt$.

On the finest level, a relaxation sweep consists of changing the coordinates 
of each path point in turn, and accepting or rejecting the new position by 
the standard Metropolis algorithm \ref{4}. 

To explain the relaxations on coarse grids, the following notation is
introduced. Let $\gij{i1}{i2}$ be the set of path points corresponding 
to times $t_{i1}, t_{i1+1}, \ldots, t_{i2}$. Let $\Sij{i1}{i2}$ be the 
action integral of $\gij{i1}{i2}$:
$$
     \Sij{i1}{i2}=\ekij{i1}{i2-1} + \dt \sumij{i1}{i2} \veff_i, \eqno{(3.1)}
$$
where $\veff$ is given by Eq. (2.6). 

A relaxation sweep on level $l>0$, is emulated by visiting all path points 
corresponding to that level. When the point $i_l$ is visited, simultaneous 
changes of the coordinates are made to $\gij{i_l-2^l}{i_l+2^l}$. These 
changes are given by 
$$
        {\overline \delta_i} = \left(1 - { {\abs{i_l-i}} \over 2^l}\right) 
        {\overline \xi} \delta_l \qquad i=i_l-2^l,\ldots,i_l,\ldots,i_l + 2^l
        \eqno{(3.2)}
$$
where $\delta_l$ is a constant, ${\overline \delta_i}$ is the 3D vector 
$(\delta_x, \delta_y, \delta_z)$ of changes to the space coordinates $(x,y,z)$
and ${\overline \xi}$ is the 3D vector $(\xi_x, \xi_y, \xi_z)$. Each of the 
components of ${\overline \xi}$ is a random variable with uniform distribution 
in $[-1,1]$ (See also Sec. 4.1). If $S^a$ is the action of 
$\gij{i_l-2^l}{i_l+2^l}$ before changes to coordinates are made, and $S^b$ is 
the action after the changes, the new configuration is accepted according 
to the transition probability
$$
        P_{ab} = \min(1, e^{-(S^b - S^a)}).
$$
Paths are generated by cycling on all levels, usually with V(1,1) cycles.
W-cycles or other cycles which increase the number of passes on coarse
levels relative to fine levels (i.e. higher cycle indices, \ref{5}
works as well, and are effective in reducing the decorrelation times between 
paths. However, in the ``unigrid'' approach, the work spent 
on a coarse level  is not much less than the work on a fine level. Therefore
there is not much to gain in efficiency by performing more relaxations on 
coarse levels. Note, however, that using higher cycle index with the
``unigrid'' approach is an effective research tool for studying the 
behavior of the algorithm.

Measurements of observables can be done either on the finest or on coarser 
levels. Because of the high correlations between adjacent path points
on the finest level, some work may be saved making measurements on coarser
levels. However, observables that depend on correlations between adjacent
path points, such as the the mean of squared velocity can not be measured on 
coarse levels.

\magnification=\magstep1
%
\def\ref#1{[#1]}
\def\delv{{\overline\delta}}
\def\delxi{{\overline\xi}}
\def\dt{\tau}
\beginsection 4. Numerical experiments \par

Numerical results for the hydrogen atom with the multigrid algorithm are
presented in Sec. 4.1. For comparison, results with a simplified 
staging algorithm (\ref{9-11}) for the same problem are presented 
in (Sec. 4.2).

All computations in this section are done with the effective potential
(2.6). The results of the calculations are not very sensitive to value of 
$A$ in (2.6). In all numerical experiments in this section $A=0.005$ is
always used.
\beginsection 4.1 Multigrid results \par

In Table 1, the results with V(1,1) cycles with the ``unigrid'' 
approach (Sec. 3) are described for different number of levels.
The meshsize of a level $l$ with $N_l$
points is $\dt_l = 32 / N_l$.

In a relaxation sweep on the finest level $(l=0)$ the $(x_i,y_i,z_i)$
coordinates are changed simultaneously at each path point by adding
$\overline\delta = (\delta_x, \delta_y, \delta_z)$ to the current values.
$\delv$ is calculated by $\delv = \delxi \delta_0$. $\delta_0$ is a
constant scalar optimized to yield Metropolis acceptance ratio of 
$\sim 0.5$ on that level. On coarser levels, simultaneous changes are
made to the 3 space coordinates of more than one point, as described 
in Sec.3. As with $\delta_0$, the scalar constant $\delta_l$ for $l>0$ is 
chosen so that the Metropolis acceptance ratio on that level is 
$\sim 0.5$.

The observables measured are the mean radius of the electron $<r>$,
the mean inverse of the radius $<1/r>$ and the mean kinetic energy 
$<e_k>$. $<e_k>$ is approximated by averaging 
the operator
$$
      (M/2\dt^2) \left((x_{i+1}-x_i)(x_i-x_{i-1}) + (y_{i+1}-y_i)(y_i-y_{i-1}) 
+
      (z_{i+1}-z_i)(z_i-z_{i-1})\right)
$$
where $\dt$ is the meshsize (\ref{12}). For each observable
the integrated decorrelation time ($\tau_{int}$) is listed.
$$
\vbox{
\offinterlineskip
\hrule
\halign{&\vrule#&\strut\quad#&\strut\quad#&\strut\quad#\hfil&\quad\vrule#&
  \quad#\hfil&\quad#\hfil&\quad\vrule#&
  \quad#\hfil&\quad#\hfil&\quad\vrule#&
  \quad#\hfil&\quad#\hfil&\quad\vrule#\cr
  height2pt&\omit&\omit&\omit&&\omit&\omit&&\omit&\omit&&\omit&\omit&\cr
  &\hfil$L$\hfil&\hfil$N_0$\hfil&\hfil$\tau_0$\hfil&&
  $<r>$&$\tau_{int}$&&
  $<1/r>$&$\tau_{int}$&&
  $<e_k>$&$\tau_{int}$& \cr
  height2pt&\omit&\omit&\omit&&\omit&\omit&&\omit&\omit&&\omit&\omit&\cr
\noalign{\hrule}
  height2pt&\omit&\omit&\omit&&\omit&\omit&&\omit&\omit&&\omit&\omit&\cr
  &\hfil 6&\hfil 128&0.2500&&
  1.431(7) & 10.0 && 1.130(4) &	7.0 && 0.80(1) & 3.0 &\cr
  &\hfil 7&\hfil 256&0.12500&&
  1.459(6) & 7.8 && 1.064(4) & 6.0 && 0.89(1) &	2.7 &\cr
  &\hfil 8&\hfil 512&0.062500&&
  1.489(7) & 9.7 && 1.019(3) & 7.0 && 0.92(1) & 2.6 &\cr
  &\hfil 9&\hfil 1024&0.031250&&
  1.493(4) & 7.3 && 1.009(2) & 5.4 && 0.99(1) & 2.4 &\cr
  &\hfil 10&\hfil 2048&0.015625&&
  1.498(3) & 8.0 && 1.004(3) & 5.5 && 0.98(1) & 2.3 &\cr
  height2pt&\omit&\omit&\omit&&\omit&\omit&&\omit&\omit&&\omit&\omit&\cr
\noalign{\hrule}
\noalign{\smallskip}
\multispan{13} $L$ - number of levels \hfil \cr
\noalign{\smallskip}
\multispan{13} $N_0$ - number of points on finest level \hfil \cr
\noalign{\smallskip}
\multispan{13} $\tau_0$ - meshsize of finest level \hfil \cr
\noalign{\smallskip}
\multispan{14} \hfil Table 1 - Multigrid V(1,1) cycles \hfil \cr
} }
$$

It is quite evident from the table that for each measured observable 
no significant CSD exists in the range of 128-2048 path points.

In Table 2 a typical behavior of $\delta_l$ as a function of $l$
is described. The results in the table are for the last case listed in
Table 1 ($L=10$, $N_0=2048$).
The Metropolis acceptance ratio (MAR) is listed in the last
column. For fine levels, $\delta_l$ is proportional to the square root of
the meshsize. For coarser levels with mean distance between grid points
comparable to $<r>$, $\delta_l$ becomes smaller with increased meshsize.
$$
\vbox{
\offinterlineskip
\hrule
\halign{&\vrule#&\strut\quad#&\quad#&\quad#&\quad\vrule#\cr
  &\hfil$l$\hfil&$\hfil\delta_l\hfil$&\hfil$MAR$\hfil&\cr
  \noalign{\hrule}
  & 0 & 0.125 & \hfil 0.505\hfil &\cr
  & 1 & 0.175 & \hfil 0.509\hfil &\cr
  & 2 & 0.245 & \hfil 0.513\hfil &\cr
  & 3 & 0.343 & \hfil 0.516\hfil &\cr
  & 4 & 0.480 & \hfil 0.518\hfil &\cr
  & 5 & 0.672 & \hfil 0.511\hfil &\cr
  & 6 & 0.941 & \hfil 0.479\hfil &\cr
  & 7 & 1.054 & \hfil 0.487\hfil &\cr
  & 8 & 0.922 & \hfil 0.490\hfil &\cr
  & 9 & 0.646 & \hfil 0.515\hfil &\cr
  \noalign{\hrule}
\noalign{\smallskip}
\multispan{4} \hfil Table 2 - behavior of $\delta_l$ \hfil \cr
  }
}            
$$

\beginsection 4.2 Staging algorithm results \par

As a comparison with the performance of the multigrid algorithm the results
of a staging algorithm for the same problem are described. This staging 
algorithm is similar to the algorithm used in \ref{6}, \ref{7},
and is a simplified version of the algorithm described in \ref{8}.

In a relaxation sweep, each point $i$ of the path becomes  an  end 
point of a chain of adjacent $p$ path points $i,i+1,\ldots,i+p+1$.
The  two end points $i$ and $i+p+1$ are held fixed and the coordinates 
of all the other chain points are changed simultaneously with means 
and variances of a Levy walk. The new chain configuration is then accepted 
or rejected. Few trials per chain may be repeated to increase efficiency.  

Table 3 describes the results of the staging algorithm for the hydrogen atom,
with different number of path points ($N$), chain length ($N_c$, including
the end points) and trials per chain ($N_{try}$). The meshsize $\tau$ is
$32/N$. The last three columns lists 1) the computational work per 
relaxation sweep ($W_r$). $W_r$ is proportional to the number of random 
numbers with uniform distribution in $[0,1]$ produced in a single relaxation 
sweep.  2) The Metropolis acceptance ratio ($MAR$) of chain states, and 3) The
integrated decorrelation times ($\tau_{int}$) for measurements of $<r>$.

As seen in Table 3, for large $N$, $MAR$ is approximately $0.5$ for 
$N_c = 5$ and $N_{try}=2$. Increasing these two numbers results only in 
very small decrease of $\tau_{int}$. Clearly, although the staging algorithm
is much faster than a primitive Metropolis algorithm, it does not eliminate
the CSD.

$$
\vbox{
\offinterlineskip
\halign{&\vrule#&\strut\quad#&\strut\quad#&\strut\quad#&\strut\quad#
   &\strut\quad#&\strut\quad#&\strut\quad#&\vrule#\cr
\noalign{\hrule}
   height2pt & \omit & \omit & \omit & \omit & \omit & \omit & \omit &\cr
   &\hfil $N$ \hfil&\hfil $\tau$ \hfil&\hfil $N_c$ \hfil&\hfil $N_{try} \hfil$
   &\hfil $W_r$ \hfil &\hfil $MAR$ \hfil &\hfil $\tau_{int} <r>$\hfil&\cr
\noalign{\hrule}
   height1pt & \omit & \omit & \omit & \omit & \omit & \omit & \omit &\cr
\noalign{\hrule}
   height1pt & \omit & \omit & \omit & \omit & \omit & \omit & \omit &\cr
   & 32 & 1.0 & 4 & 1 & 0.21 & 0.32 & 15.6 &\cr
   &\omit&\omit&\omit& 2 & 0.44 & 0.67 & 7.7 &\cr
   &\omit&\omit& 5 & 2 & 0.64 & 0.32 & 7.8 &\cr
   &\omit&\omit&\omit& 3 & 0.95 & 0.48 & 7.7 &\cr
   &\omit&\omit& 6 & 4 & 1.7 & 0.24 & 9.5 &\cr
   &\omit&\omit&\omit& 5 & 2.1 & 0.32 & 8.5&\cr
   height1pt & \omit & \omit & \omit & \omit & \omit & \omit & \omit &\cr
\noalign{\hrule}
   height1pt & \omit & \omit & \omit & \omit & \omit & \omit & \omit &\cr
   & 64 & 0.5 & 4 & 1 & 0.44 & 0.45 & 33.0 &\cr
   &\omit&\omit& 5 & 2 & 1.3 & 0.46 & 14.6 &\cr
   &\omit&\omit&\omit& 3 & 1.9 & 0.67 &\omit&\cr
   &\omit&\omit& 6 & 3 & 2.5 & 0.31 &\omit&\cr
   &\omit&\omit&\omit& 4 & 3.3 & 0.41 &\omit&\cr
   &\omit&\omit&\omit& 5 & 4.1 & 0.50 & 12.5 &\cr
   height1pt & \omit & \omit & \omit & \omit & \omit & \omit & \omit &\cr
\noalign{\hrule}
   height1pt & \omit & \omit & \omit & \omit & \omit & \omit & \omit &\cr
   & 128 & 0.25 & 4 & 1 & 0.86 & 0.50 & 120.0 &\cr
   &\omit&\omit& 5 & 1 & 1.30 & 0.25 & \omit & \cr
   &\omit&\omit&\omit & 2 & 2.5 & 0.51 & 75.0 & \cr 
   height1pt & \omit & \omit & \omit & \omit & \omit & \omit & \omit &\cr
\noalign{\hrule}
   height1pt & \omit & \omit & \omit & \omit & \omit & \omit & \omit &\cr
   & 256 & 0.125  & 4 & 1 & 1.7 & 0.51 &\omit &\cr
   &\omit & \omit & 5 & 1 & 2.5 & 0.27 &\omit &\cr
   &\omit & \omit & \omit& 2 & 5.1 & 0.53 & 350.0 &\cr
   height1pt & \omit & \omit & \omit & \omit & \omit & \omit & \omit &\cr
\noalign{\hrule}
   height1pt & \omit & \omit & \omit & \omit & \omit & \omit & \omit &\cr
   & 512 & 0.0625 & 5 & 2 & 10.0 & 0.53 & 1240.0 &\cr
   height1pt & \omit & \omit & \omit & \omit & \omit & \omit & \omit &\cr
\noalign{\hrule}
\noalign{\medskip}
\multispan{8} \hfil Table 3 - Staging algorithm \hfil \cr
} }
$$

\vfill

\beginsection Acknowledgements \par
The author with to thank A. Brandt for helpful discussions and M.H. Kalos 
for his helpful suggestions.

\bigskip
\centerline{\hfil Bibliography \hfil}
\medskip
\item{[\ 1]} W. Janke and T. Sauer {\it Path integral Monte Carlo using
multigrid techniques }, Chemical Review Letters, {\bf 201} 5,6, 499 (1993). 
\item{[\ 2]} M. H. Kalos {\it Stochastic wave function for atomic helium},
J. Comput. Phys. {\bf 2}:257 (1967).
\item{[\ 3]} S. Zhang and M. H. Kalos {\it Bilinear Monte Carlo: Expectations
and energy differences}, J. Stat. Phys. {\bf 70}:515 (1993).
\item{[\ 4]} N. Metropolis, A. Rosenbluth, M.N. Rosenbluth, A.H. Teller
and E. Teller {\it Equation of state computing by fast computing machines},
J. Chem. Phys. {\bf 21} 6, pp. 1087-1092 (1953).
\item{[\ 5]} A. Brandt, M.Galun and D. Ron {\it Optimal multigrid algorithms for
calculating thermodynamic limits}, availabe from Department of Applied
Mathematics, Weizmann Institute of Science, Rehovot, Israel. 
\item{[\ 6]} D.F. Coker, B.J. Berne and D. Thirumalai {\it Path integral Monte
Carlo studies of the behavior of excess electrons in simple fluids} J. Chem.
Phys. {\bf 86}(10) pp. 5689-5702 (1987).
\item{[\ 7]} Jianshu Cao and Bruce J. Berne {\it On energy estimators in path
integral Monte Carlo simulations: Dependence of accuracy on algorithm} J.
Chem. Phys. {\bf 91}(10) pp. 6359-6366 (1989).
\item{[\ 8]} E.L. Pollok and D.M. Ceperley {\it Simulation of quantum many-body
systems by path-integral methods} Phys. Rev. B {\bf 30}, 2555 (1984).
\item{[\ 9]} M. Sprik, M.L. Klein, and D. Chandler, Phys. Rev. B {\bf 32},
545 (1985).
\item{[10]} M. Sprik, M.L. Klein, and D. Chandler, Phys. Rev. B {\bf 81},
4234 (1985).
\item{[11]} M. Sprik, M.L. Klein, and D. Chandler, J. Chem. Phys, {\bf 83},
3042 (1985).
\item{[12]} R. P. Feynman and A.R. Hibbs, ``{\it Quantum mechanics and path
integrals}'', McGraw-Hill, New-York 1965. 

\vfill
\bye